\documentclass{PoS}

\usepackage[utf8]{inputenc}
\usepackage{amsmath}
\usepackage{amssymb}
\usepackage{color}
\usepackage{graphicx}
\usepackage{multirow}

\usepackage{subcaption}
\usepackage{braket}

\title{Lattice QCD study of heavy-heavy-light-light tetraquark candidates}

\ShortTitle{Lattice QCD study of heavy-heavy-light-light tetraquark candidates}

\author{\speaker{Antje Peters}$^a$, Pedro Bicudo$^b$, Luka Leskovec$^{c}$, Stefan Meinel$^{c,d}$, Marc Wagner$^a$ \\

$^a$Goethe-Universit\"at Frankfurt am Main, Institut f\"ur Theoretische Physik,\\
Max-von-Laue-Stra{\ss}e 1, D-60438 Frankfurt am Main, Germany\\

$^b$CFTP, Departamento de Física, Instituto Superior Técnico, Universidade de Lisboa,\\
Avenida Rovisco Pais, 1049-001 Lisboa, Portugal\\

$^c$Department of Physics, University of Arizona,\\
1118 E 4th Street, Tucson, AZ 85721, USA\\

$^d$RIKEN BNL Research Center, Brookhaven National Laboratory,\\
 PO Box 5000,  Upton, NY 11973, USA\\

E-mail: \email{peters@th.physik.uni-frankfurt.de, bicudo@tecnico.ulisboa.pt, leskovec@email.arizona.edu, smeinel@email.arizona.edu, mwagner@th.physik.uni-frankfurt.de}
}

\abstract{
 We investigate heavy-light four-quark systems $ud\bar b \bar b$ with bottom quarks of finite mass which are treated in the framework of NRQCD. We focus on $I(J^P)=0(1^+)$, where we recently found evidence for the existence of a tetraquark state using static bottom quarks. Furthermore, we report on an investigation of the $u \bar d b \bar b$  four-quark system with quantum numbers $I(J^P)=1(1^+)$ again using static bottom quarks.
}

\FullConference{The 34th International Symposium on Lattice Field Theory\\
                 24-30 July  2016\\
                 Highfield Campus, University of Southampton, Southampton, UK}

\begin{document}


\section{Motivation}

In the near future, heavy-light tetraquarks are expected to be studied in more detail by experi\-mental collaborations. Therefore, it is crucial to gain deeper theoretical understanding of heavy-light four-quark systems. On the one hand, the nature of already observed four-quark states has to be investigated. Two examples for tetraquark candidates are the charged bottomonium-like and charmonium-like mesons $Z_b^\pm$ and $Z_c^\pm$, respectively (cf.\ e.g.\ \cite{Belle:2011aa}). On the other hand, it is important to predict possibly existing four-quark systems to give directions for future experimental research.

Here we report on our investigations of four-quark systems with two heavy (anti-)quarks $\bar b \bar b$ and $b \bar b$ and two lighter quarks
$qq$ and $q\bar  q$, $q\in \{u,d\}$. 

Previous papers with similar studies of $ud\bar b\bar b$ and $u\bar d b\bar b$ systems include\ \cite{Wagner:2010ad,Wagner:2011ev,Bicudo:2012qt, Wagenbach:2014oxa, Bicudo:2015vta, Scheunert:2015pqa, Peters:2015tra, Bicudo:2015kna, Stewart:1998hk,Michael:1999nq,Cook:2002am,Doi:2006kx,Detmold:2007wk,
Bali:2010xa,Brown:2012tm, Francis:2016hui,Alberti:2016dru}.


\section{$ud\bar b\bar b$ systems with non-relativistic bottom quarks}

In a recent study of $ud\bar b \bar b$ four-quark systems using static bottom quarks \cite{Bicudo:2012qt,Bicudo:2015vta,Bicudo:2015kna} we found evidence for a bound state with corresponding quantum numbers $I(J^P)=0(1^+)$. The binding energy of this state is $E_{\textrm{binding}}=90^{+36}_{-43}$ MeV. In another recently published study \cite{Francis:2016hui} using bottom quarks of finite mass similar results were obtained. In this work we again investigate $ud\bar b\bar b$ systems. This time we also use bottom quarks of finite mass by means of NRQCD.

\subsection{Lattice QCD setup}

The positions of bottom quarks of finite mass are not fixed. Therefore, in contrast to our previous study \cite{Bicudo:2012qt,Bicudo:2015vta,Bicudo:2015kna}, the computation of a $\bar b \bar b$ potential in the presence of two light quarks $ud$ is not possible. However, one can directly compute the energy of the lowest $ud\bar b \bar b$ state in the $I(J^P)=0(1^+)$ channel. If it is a bound four-quark state, this energy $E_{ud\bar b \bar b}$ should be below the $B B^\ast$ threshold, i.e.\ $E_{ud\bar b \bar b} < E_B + E_{B^\ast}$. The structure of such a state could resemble two loosely bound mesons, i.e.\ $B B^*$ or $B^\ast B^\ast$, or a diquark-antidiquark pair.

In a first step we have computed the effective energy of a $B B^\ast$ molecule-like operator as well as the effective energies of standard $B$ and the $B^*$ operators,
\begin{equation}
aE^{(\textrm {eff})}_{i}(t)=\ln\Big(\mathcal C_{i}(t)/\mathcal C_{i}(t+a)\Big) \quad , \quad i = B B^\ast\textrm{-mol} , B , B^\ast ,
\label{effm}
\end{equation}
where
\begin{eqnarray}
\nonumber & & \hspace{-0.7cm} \mathcal C_{B B^\ast{\scriptsize \textrm{-mol}}}(t)=\sum\limits_{ \vec x^\prime} 
\Big \langle \Big( \bar b(\vec x,0)\Gamma_1 d(\vec x,0) \bar b(\vec x,0) \Gamma_2 u(\vec x,0) -  \bar b(\vec x,0)\Gamma_1 u(\vec x,0) \bar b (\vec x,0)\Gamma_2 d(\vec x,0)\Big) \\
 & & \hspace{1.5cm} \Big( \bar d(\vec x^\prime,t)\Gamma_1^\prime b(\vec x^\prime, t) \bar u(\vec x^\prime,t) \Gamma_2^\prime b(\vec x^\prime,t) - \bar u(\vec x^\prime,t) \Gamma_1^\prime b(\vec x^\prime,t) \bar d(\vec x^\prime,t) \Gamma_2^\prime b(\vec x^\prime,t)\Big)\Big \rangle \\
 & & \hspace{-0.7cm} \mathcal C_{B}(t)=\sum\limits_{ \vec x^\prime}
\Big \langle \bar b(\vec x,0)  \Gamma_1 d(\vec x,0)  \bar d(\vec x^\prime,t) \Gamma_1^\prime b(\vec x^\prime,t) \Big \rangle \\
 & & \hspace{-0.7cm} \mathcal C_{B	^*}(t)=\sum\limits_{ \vec x^\prime}  \Big \langle \bar b(\vec x,0)\Gamma_2 d(\vec x,0)
 \bar d(\vec x^\prime,t) \Gamma_2^\prime b(\vec x^\prime,t) \Big \rangle
\end{eqnarray} 
and $\Gamma_1=\gamma_5 $, $\Gamma_2=\gamma_i $, $\Gamma_1^\prime=-\gamma_5$ and $\Gamma_2^\prime=-\gamma_i^\dagger$.
We use gauge-covariant Gaussian smearing for all quark fields. 

Computations have been performed using gauge link configurations generated by the RBC and UKQCD collaborations with  the Iwasaki gauge action and $n_f=2+1$ domain-wall fermions. Information on these configurations can be found in Table \ref{confsNRQCD} or in \cite{Brown:2014ena}.

\begin{table}[hb]
	\begin{tabular}{ccccccccc}
  	  Ens. & $\beta$ & lattice & $a m_{u,d}$ & $a m_{s}$ & $m_{\pi}$[MeV] & $a$ [fm] & $L$ [fm] & measurements\\
		\hline
 	 \texttt{C54}   & 2.13 & $24^3\times 64$  & 0.005 &0.04& 336 & 0.1119(17) &  2.7  & 1676 \\
  \end{tabular}
  \caption{Ensemble of gauge link configurations. $\beta$: inverse gauge coupling, $m_{u,d}$: $u/d$ light sea and valence quark mass, $m_{s}$:  strange sea and valence quark mass, $m_\pi$: pion mass,  $a$: lattice spacing, $L$: lattice extent in fm, measurements: number of samples taken on different gauge link configurations or different source locations.}
  \label{confsNRQCD}
\end{table}

\begin{figure}[htb]
\begin{center}
\includegraphics[width=10.0cm]{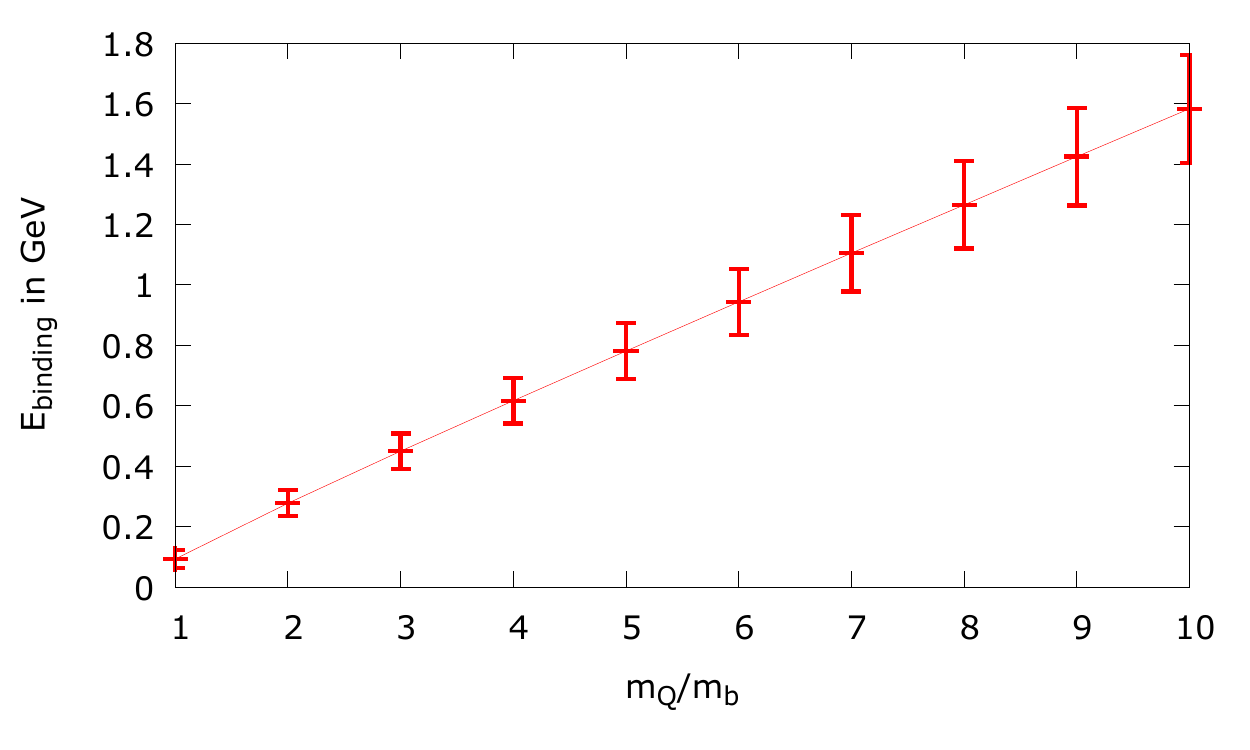}
\caption{The binding energy $E_{\textrm{binding}}$ of a four-quark state from static-light potential calculations, as a function of the heavy quark mass $m_Q$ in units of the physical mass $m_b$.}
\label{unphysicalmb}
\end{center}
\end{figure}

The bottom quark is treated in the framework of NRQCD and the action is tadpole-improved (cf. \cite{Brown:2014ena} for further details). We have performed computations with a bottom quark mass $m_Q$ corresponding to the physical value $m_b$ as well as to $5 m_b$. We decided to also study such unphysically heavy bottom quarks, because from our previous static-light computations \cite{Bicudo:2012qt,Bicudo:2015vta,Bicudo:2015kna} we expect a clear signal for a four-quark bound state with a large binding energy ($\approx 800$ MeV) (cf. Figure \ref{unphysicalmb}).

\subsection{Numerical results}

\begin{figure}[htb]
\begin{center}
\includegraphics[width=10.0cm]{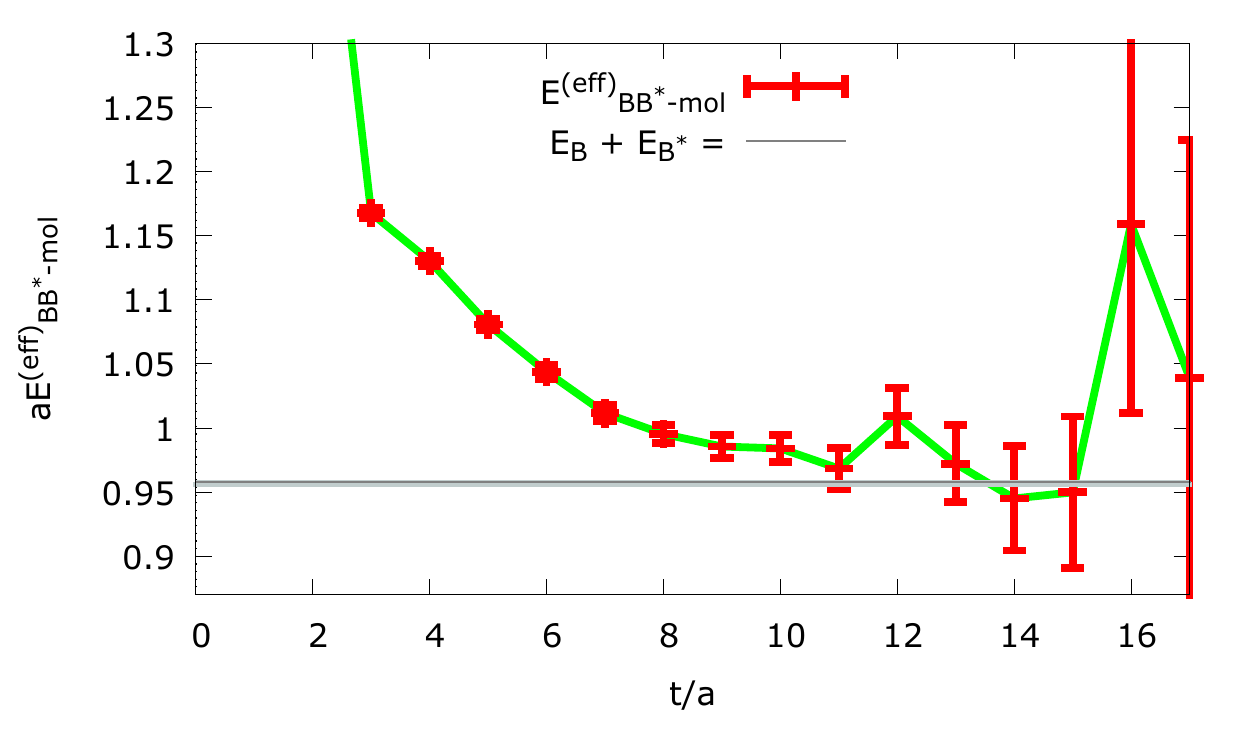} \\
\includegraphics[width=10.0cm]{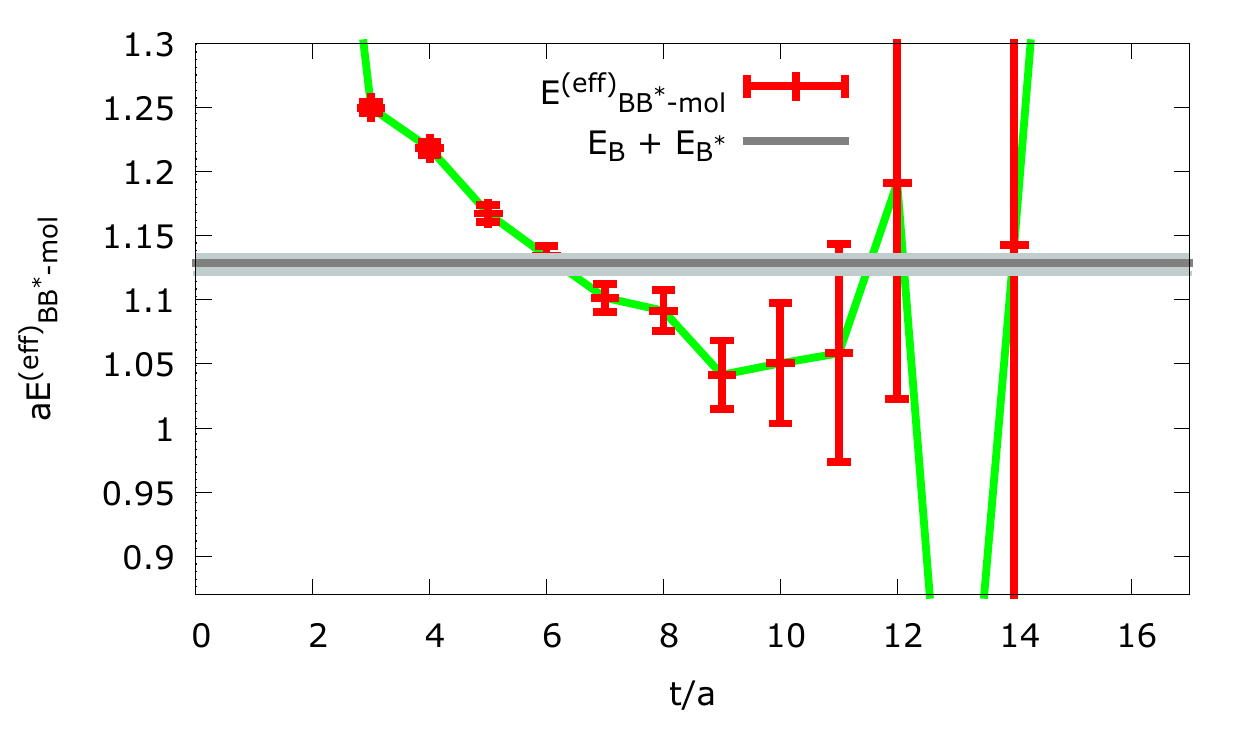}
\end{center}
\caption{The effective energy corresponding to the $B B^\ast$ molecule-like operator in units of the lattice spacing as a function of the temporal separation $t/a$ (red points) in comparison to $E_B+E_{B^*}$ (horizontal grey line). \textbf{(top)}: $m_Q=m_b$. \textbf{(bottom)}: $m_Q=5 m_b$.}
\label{meffplot}
\end{figure}

The effective energies $E^{(\textrm {eff})}_{B B^\ast\textrm{-mol}}$ corresponding to the $B B^\ast$ molecule-like operator are shown in Figure~\ref{meffplot} both for $m_Q=m_b$ and $m_Q=5m_b$. The horizontal grey lines and error bands correspond to the sum of the energies of the mesons $E_B + E_{B^*}$, which can be obtained rather precisely by fitting constants to the corresponding effective energies (\ref{effm}) at large temporal separations $t$.

Even though for $m_Q = 5 m_b$ the effective energy seems to be below $E_B + E_{B^*}$ for large temporal separations $t$, which might be an indication for a bound four-quark state, the plateau quality is at the moment not sufficient to make solid statements. Besides increasing the statistical accuracy, we plan to include additional four-quark operators in our study. In particular, we intend to include $B^\ast B^\ast$ molecule-like, diquark-antidiquark and $B B^\ast$ scattering operators. Such an extended set of operators should allow to disentangle any mixing of low-lying states, which we might currently observe. Moreover, possibly significant finite-volume effects need to be excluded or taken into account.

\section{The $u\bar d b \bar b$ system with static bottom quarks}

A frequently discussed and investigated tetraquark candidate is the recently measured $Z_b^+$ meson with quantum numbers $I(J^P)=1(1^+)$.

The isospin $I=1$ can be realized by light quark flavours $u\bar d$. Moreover, a loosely bound $B \bar B^*$ system can account for spin $J=1$ and parity $P=+$, since $B$ has quantum numbers $J^P=0^-$ and $\bar B^*$ has quantum numbers $J^P=1^-$. However, the same quantum numbers can also be realized by other structures. One possibility is a bottomonium-like state $b \bar b$ in combination with a far separated pion $\pi^+$. A technical challenge regarding the $u\bar d b \bar b$ system with $I(J^P)=1(1^+)$ is to disentangle those structures to find the correct binding energy of a possibly existing tetraquark.

To determine the binding energy of the $B\bar B$ state, we use static bottom quarks and the Born-Oppenenheimer approximation \cite{BO}, i.e.\ we proceed in two steps.  The first step  is the computation of the potential of two static quarks in the presence of two quarks of finite mass by means of lattice QCD. In case of static (anti-)quarks $b$ and $\bar b$ and light (anti-)quarks $u$ and $\bar d$, the potential at large separations between $b$ and $\bar b$ can be interpreted as as the potential between a $B$ and a $\bar B^*$ meson. The second step is to solve Schrödinger's equation to check whether the respective potential is sufficiently attractive to host a bound state.

We find a binding energy of 
\begin{equation*}
E_{\textrm{binding}} =(58 \pm 71) \, \textrm{MeV} .
\end{equation*}
which is a very vague signal for the existence of an $I(J^P)=1(1^+)$ $ u\bar d b\bar b$ tetraquark. Our findings are, hence, consistent with the experimentally observed $Z_b^\pm$ states. For more details regarding this investigation cf.\ \cite{Peters:2016wjm}.


\section{Summary and outlook}

$ud \bar b \bar b$ systems are experimentally hard to observe, but theoretically straightforward to investigate. In a previous study \cite{Bicudo:2015vta, Peters:2015tra, Bicudo:2015kna} we found evidence for a bound $ud\bar b \bar b$ state with quantum numbers $I(J^ P)=0(1^ +)$ in the static-light approximation. Using NRQCD we are now in the process of studying these systems with bottom quarks of finite mass.

$u\bar d b \bar b$ systems are experimentally easier to access than $ud \bar b \bar b$ systems, but theoretically more challenging. We are currently investigating $I(J^ P)=1(1^+)$ corresponding to $Z_b^+$ in the static-light approximation. We we find a binding energy $E_{\textrm{binding}}=(58 \pm 71)$ MeV.


\acknowledgments

P.B.\ thanks IFT for hospitality and CFTP, grant FCT UID/FIS/00777/2013, for support. S.M.\ is supported by National Science Foundation Grant Number PHY-1520996, and by the RHIC Physics Fellow Program of the RIKEN BNL Research Center. M.W.\ and A.P.\ acknowledge support by the Emmy Noether Programme of the DFG (German Research Foundation), grant WA 3000/1-1. 

This work was supported in part by the Helmholtz International Center for FAIR within the framework of the LOEWE program launched by the State of Hesse.

Calculations on the LOEWE-CSC high-performance computer of Johann Wolfgang Goethe-University Frankfurt am Main were conducted for this research. We would like to thank HPC-Hessen, funded by the State Ministry of Higher Education, Research and the Arts, for programming advice.



\end{document}